# CMOS LOW POWER CELL LIBRARY FOR DIGITAL DESIGN

Kanika Kaur[1], Arti Noor[2]

Research Scholar, JJTU, Rajasthan[1], CDAC, Noida, U.P[2]
kanika.kiit@gmail.com

**ABSTRACT**

*Historically, VLSI designers have focused on increasing the speed and reducing the area of digital systems. However, the evolution of portable systems and advanced Deep Sub-Micron fabrication technologies have brought power dissipation as another critical design factor. Low power design reduces cooling cost and increases reliability especially for high density systems. Moreover, it reduces the weight and size of portable devices. The power dissipation in CMOS circuits consists of static and dynamic components. Since dynamic power is proportional to $V^2_{dd}$ and static power is proportional to $V_{dd}$, lowering the supply voltage and device dimensions, the transistor threshold voltage also has to be scaled down to achieve the required performance.*

*In case of static power, the power is consumed during the steady state condition i.e when there are no input/output transitions. Static power has two sources: DC power and Leakage power. Consecutively to facilitate voltage scaling without disturbing the performance, threshold voltage has to be minimized. Furthermore it leads to better noise margins and helps to avoid the hot carrier effects in short channel devices. In this paper we have been proposed the new CMOS library for the complex digital design using scaling the supply voltage and device dimensions and also suggest the methods to control the leakage current to obtain the minimum power dissipation at optimum value of supply voltage and transistor threshold. In this paper CMOS Cell library has been implemented using TSMC (0.18um) and TSMC (90nm) technology using HEP2 tool of IC designing from Mentor Graphics for various analysis and simulations.*

## 1. INTRODUCTION

In the earlier period, the chief concern of the VLSI designer was area, performance, cost and reliability; power considerations were mostly of only secondary significance. The benefit of utilizing an arrangement of low-power components in conjunction with low-power design techniques is more important now than ever before. Necessities for lower power consumption continue to increase significantly as components become battery-powered, smaller and needed more functionality [16]. In the earlier period the main concerns for the VLSI designers was area, performance and cost. Power consideration was the secondary concerned. Now a day's power is the prime concerned due to the significant growth and achievement in the field of portable, compact & personal computing devices and wireless communication system which demand high speed computation and complex functionality with low power consumption [16].

In this paper we have proposed power reduction techniques at circuit level using voltage scaling i.e by optimizing Vdd. At lower Vdd the subthreshold current is increased, to overcome this **tox** and Capacitance can reduce. We have implemented a CMOS cell library for digital circuit designs at (0.18 micron and 90 nm) technology using voltage scaling and controlling the subthreshold leakage current without affecting the performance. The techniques are based on the fact that, under inversion bias, gate leakage through SiO2 for the PMOS transistors is an order of magnitude lower than for the NMOS [3].The rest of the paper is organized as follows: Section 2 describes the sources of power Dissipation Section 3 describes the various techniques of power minimization. Section 4 describes the proposed CMOS cell library at various technology and comparison between the various analyses. Section 5 describes the result and finally, section 6 concludes the paper.

## 2. SOURCES OF POWER DISSIPATION

CMOS is, by far, the most common technology used for manufacturing digital ICs. There are 3 major sources of power dissipation in a CMOS circuit [9]:

**P = PSwitching + PShort-Circuit + PLeakage- (1)**

PSwitching, known as switching power, is due to discharging and charging capacitors driven by the circuit [9]. PShort-Circuit is known as short-circuit power, is generated by the short circuit currents that occur when pairs of PMOS/NMOS transistors are conducting concurrently [9]. PLeakage is known as leakage power, generate from substrate injection and subthreshold effects. For older technologies, PSwitching was predominant. For deep-submicron processes, PLeakage becomes more important. Design for low-power involves the ability to minimize all three components of power consumption in CMOS circuits during the implementation and growth of a low power electronic product. The development of process scaling for CMOS technology has made subthreshold leakage reduction a growing issue for submicron circuit designers as[7] in fig.(1).

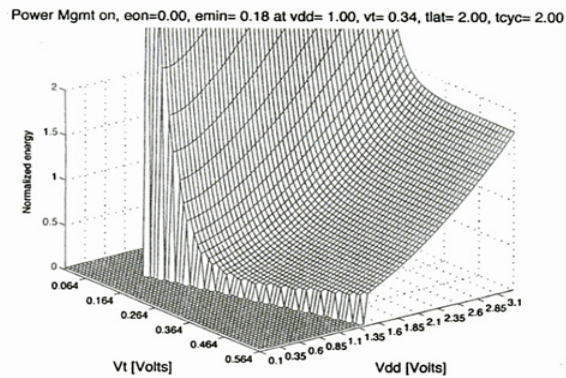

Fig1.Power dissipation as function of supply voltage (Vdd) and Threshold voltage (Vth)[7]

## 3. POWER MINIMIZATION TECHNIQUES

### 3.1 Voltage Scaling

Voltage scaling is perhaps the most effective method of saving power due to the square law dependency of digital circuit active power on the supply voltage [10].

For a microelectronics device's total power consumption can be represented by:

$$Ptot = \alpha C_{tot} V_{dd}^2 f + V_{dd} I_{off} \quad \text{------------- (2)}$$

The first term in Equation 2 represents dynamic or "switching" power, while the second term represents static power which happens due to the leakage in the design. VDD Scaling is the preferable method implemented for low power design but it decreases the circuit speed since VGS – VT, is reduced. To deal with this, systems may utilize dynamic voltage scaling to permit the lowest VDD essential to meet the circuit speed necessities while saving the energy used for the computation [9]- [12]. Supply voltage scaling enhances the gate delays unless the threshold voltage of the transistors is also scaled down. Due to the minimization of the threshold voltage there is a considerable rise in the leakage current of the transistors.

Hence, there is a clear tradeoff among the active power and off-state leakage for a specified application, leading to methodical selection of VDD and VT for performing an assigned task [6].

### 3.2 Reducing the physical Capacitance

Digital circuits have three types of capacitance: gate capacitance, diffusion capacitance and interconnect capacitance. If all the three components are scaled down as well by the same factor, then the net power dissipation is scaled down as well[17]. Gate and diffusion capacitance are fixed during the cell design, whereas Intercell and global interconnect capacitances can be controlled by the CAD tools performing the global routing [17]. Physical capacitance mainly reduces by the transistor sizing [8].

### 3.3 Reducing the switching frequency

Reducing the number of "0"to '1' power dissipating transitions minimize the switching power dissipation of the gate .Switching frequency may be reduced on several levels in the design process beginning from circuit level to the architectural level [17]. There are several logic styles to design with. Some of these styles are: Static CMOS, CPL, MCML, and a variety of dynamic logic styles. Generally, most logic styles perform delay power tradeoffs, but not always in proportional amounts [17]. The best style is that which minimize power dissipation given a constant throughput [17].

## 4. PROPOSED CMOS CELL LIBRARY OF LOGIC GATES

The proposed CMOS cell library of logic gates was designed using HEP2 of Mentor Graphics Tool at 130nm and 90nm technology with supply voltage of 1.2V. We have designed all the logic gates, basic combinational and sequential circuits for digital design based on voltage scaling and leakage current and ISub can be minimize using Stacking/ MVL method.

International Journal of VLSI design & Communication Systems (VLSICS) Vol.4, No.3, June 2013

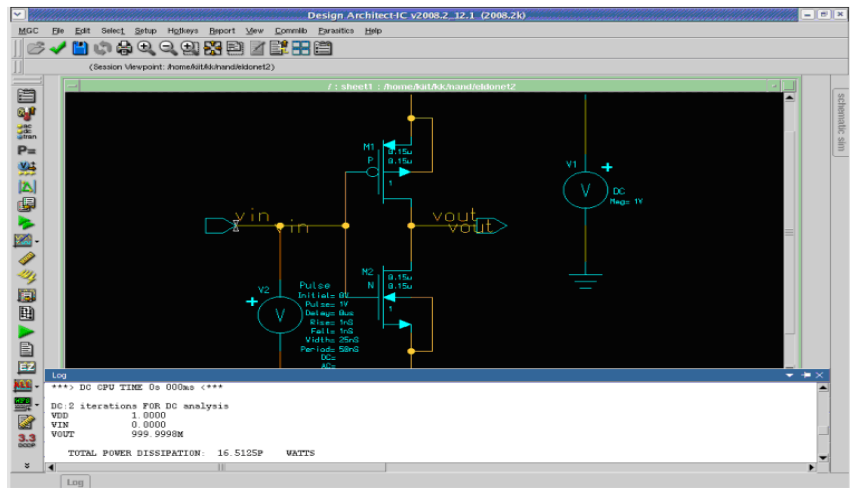

Fig.2 CMOS Cell of Not gate at 130 nm technology

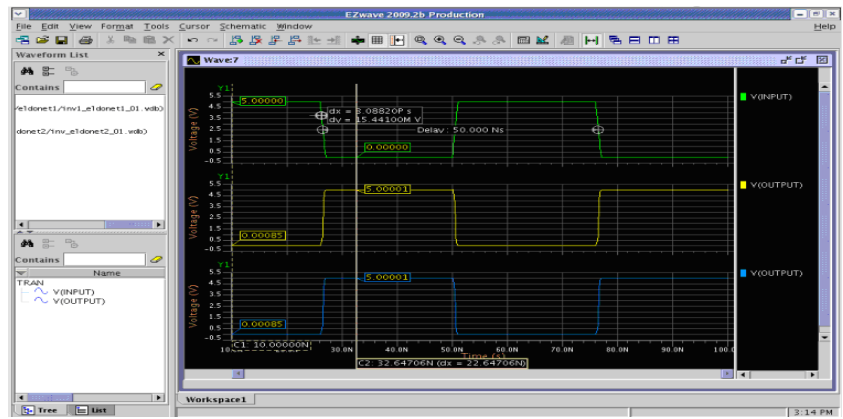

Fig.3 O/P of CMOS Cell of Not gate at 130 nm technology

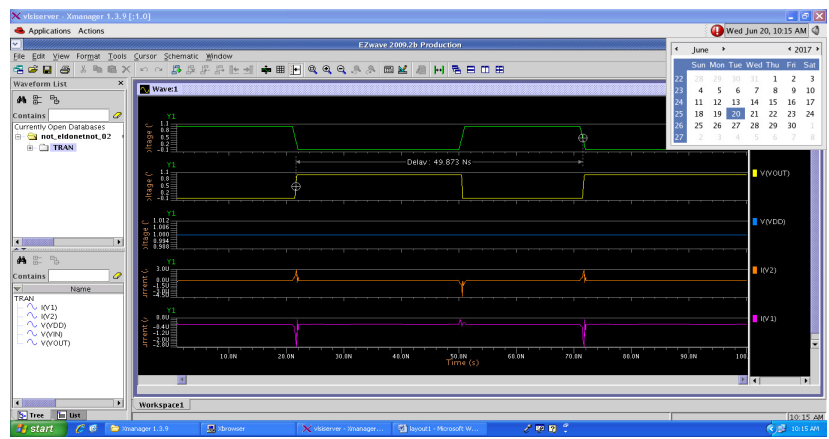

Fig4. O/P of CMOS Cell of Not gate at 90 nm technology

46

International Journal of VLSI design & Communication Systems (VLSICS) Vol.4, No.3, June 2013

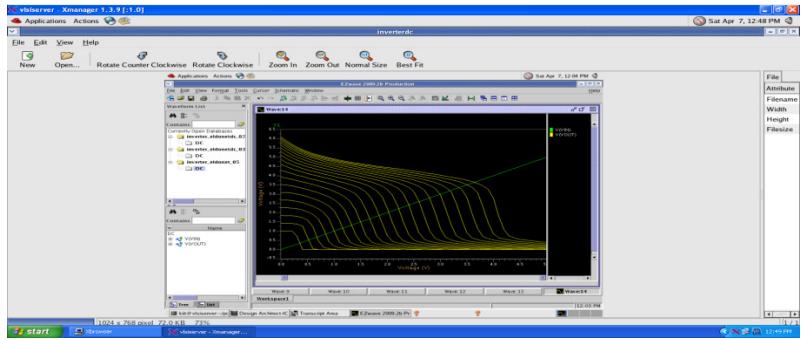

Fig5. Comparison of Isub at multi Vdd

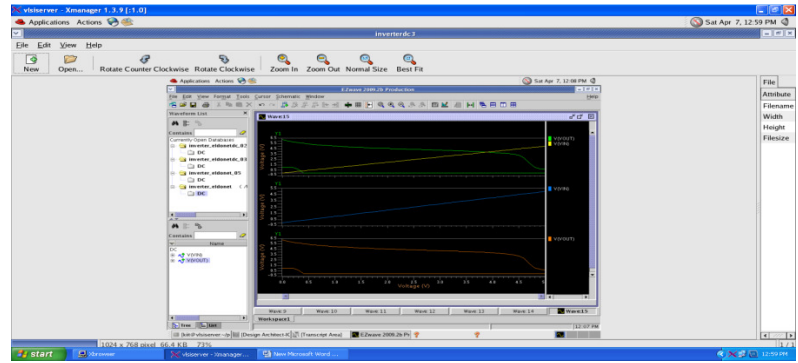

Fig.6 Comparison of leakage current at Subthreshold and Superthreshold region

## 5. RESULTS AND DISCUSSIONS

From the table 2.1, 2.2, & 2.3 we can easily compare the power dissipation, delay, leakage power and area of three logic cells. The performance of the logic cells were verified under Process corner analysis where propagation delay, leakage power and power dissipation were calculated compares it with conventional method, load analysis and stacking method. It is clear from the table that there is the reduction in leakage power for each cell implemented using stacking method than using conventional method. But stacking method results in increase of delays and routing problems as compared to conventional method.

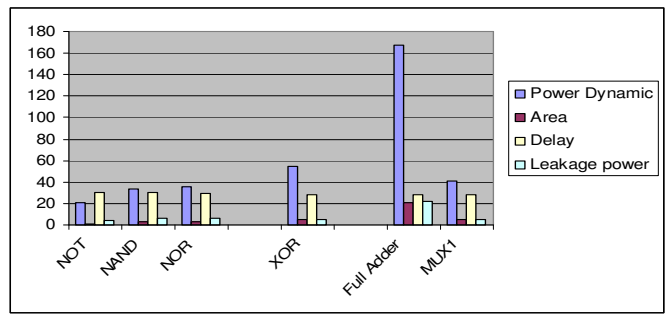

Fig.7 Performances of Conventional Circuits

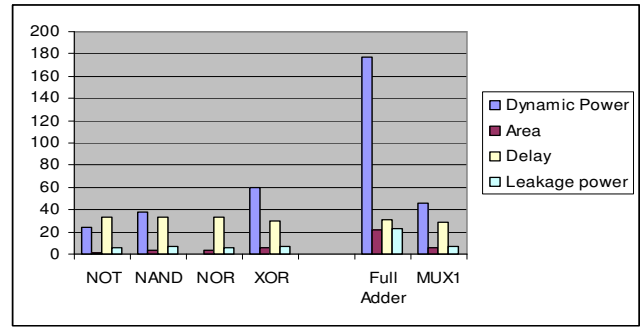

Fig.8 Performances of Conventional Circuits with load

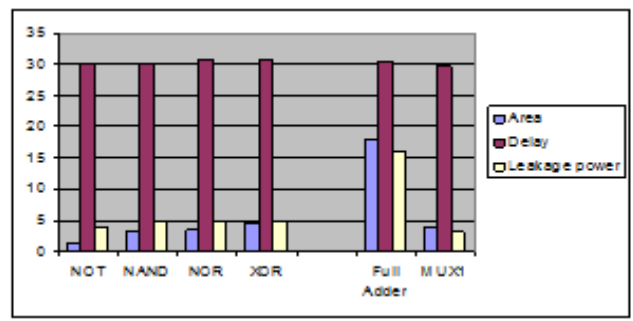

Fig. 9 Performance of Circuit with Stacking

## 6. CONCLUSION

This paper is focused on Low power VLSI cell library for the digital design at transistor gate level. This paper describes the low power techniques voltage scaling and stacking effect.

The performance of the logic cells were verified under Process corner analysis where propagation delay, leakage power and power dissipation were calculated compares it with conventional method, load analysis and stacking method and these cells were implemented on FPGA Board. Thus this technique can be used to design and characterize a new cell library for ultra low power cells in deep sub-micron region and which can work at RF Level.

## Author

Kanika Kaur (Associate Professor, KIIT, Gurgaon) received B.Sc (Electronics) Hons. Degree from Delhi University in 1997 and M.Sc Electronics) Hons. Degree from Jamia Millia Islamia University in 1999.She received M.Tech degree from RTU in 2005 and presently pursuing Ph.D from the JJTU, Rajasthan in the field of "Low power VLSI design-subthreshold leakage reduction technique for CMOS". Published more than 20 research papers in national, international journal & conferences. She has also published a book titled "Digital System Design" by SciTech Publication in 2009.Editor of 05 Technical Proceedings of National & International Seminars. Convener of many National and International Symposium. Life member of IETE & ISTE. Awarded as best academic personality& HOD in 2007 and 2008 NIEC, Delhi. Convener of Research Journal of KIIT College of Engineering. Gurgaon.

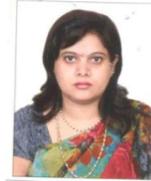

Tables

Table 2.1 performance of conventional circuit

| Logic Cell | Area (Sq.um) | Delay (50%) | Leakage power (nW) |
|---|---|---|---|
| NOT | 1.32 | 30.327ns | 3.98 |
| NAND | 3.322 | 30.339ns | 5.00 |
| Full Adder | 17.89 | S=30.456ns C=30.768ns | 16.02 |
| MUX1 | 3.97 | 29.635ns | 3.15 |

Table 2.2   Load analysis of cells

| Logic Cell | Power Dissipation (Active/Dynamic) | Area (Sq.um) | Delay (50%) | Leakage power (nW) |
|---|---|---|---|---|
| NOT | 20.801pw | 1.32 | 29.873ns | 4.27 |
| NAND | 33.782pw | 3.322 | 29.853ns | 5.87 |
| Full Adder | 167.2793pw | 20.85 | S=28.762ns C=28.666ns | 21.91 |
| MUX1 | 40.943pw | 4.93 | 28.535ns | 5.28 |

Table 2.3, Stacking Method for leakage control

| Logic Cell | Power Dissipation (Active/Dynamic) | Area (Sq.um) | Delay (50%) | Leakage power (nW) |
|---|---|---|---|---|
| NOT | 23.6805pw | 1.56 | 32.873ns | 5.75 |
| NAND | 37.456pw | 3.584 | 32.853ns | 6.79 |
| Full Adder | 176.880pw | 21.98 | S=30.62 ns<br>C=30.67ns | 23.08 |
| MUX1 | 45.894pw | 5.64 | 28.535ns | 6.99 |